\documentclass[aip, amsmath, amssymb, reprint, a4paper]{revtex4-2}
\usepackage{graphicx}
\usepackage{hyperref}
\usepackage{natbib}
\usepackage{physics}
\usepackage{xcolor}

\begin{document}
\title{Analysing the Grain size and asymmetry of the particle distribution using auto-correlation technique}

\author{Vanitha Patnala} 
\affiliation{SRM University -- AP, Amaravati, Mangalagiri, Andhra Pradesh, 522502, India.}

\author{Salla Gangi Reddy} \email{gangireddy.s@srmap.edu.in}
\affiliation{SRM University -- AP, Amaravati, Mangalagiri, Andhra Pradesh, 522502, India.}

\author{Shashi Prabhakar}
\affiliation{Quantum Science and Technology Laboratory, Physical Research Laboratory, Navrangpura, Ahmedabad 380009, India.}

\author{R. P. Singh} \email{rpsingh@prl.res.in}
\affiliation{Quantum Science and Technology Laboratory, Physical Research Laboratory, Navrangpura, Ahmedabad 380009, India.}

\author{Venkateswarlu Annapureddy} \email{annp@nit.edu}
\affiliation{Department of Physics, National Institute of Technology, Tiruchirappalli, Tamil Nadu 620015, India}

\begin{abstract}
Extracting the grain size from the microscopic images is a rigorous task involving much human expertise and manual effort. While calculating the grain size, we will be utilizing a finite number of particles which may lead to an uncertainty in the measurement. To avoid this difficulty, we utilize a simple mathematical tool, the auto-correlation function, to determine the grain size. The random particle distribution and the finite width Gaussian histogram of particle size has motivated us to utilize the auto-correlation function, which has been extensively studied for finding the size of random optical patterns. The finite width of the correlation function provides the grain size, and the difference in correlation length along two mutually independent directions provides information about the asymmetry present in the particle distribution, i.e., the deviation from a spherical shape. The results may find applications in material, pharmaceutical, chemical, and biological studies where extracting the grain size is essential. 
\end{abstract}

\maketitle

It is known that the grain size plays a vital role in controlling the electrical, magnetic, magneto-electric, mechanical and optical properties of the materials \cite{yuan2014dependence, ponpandian2002influence, matsumoto2002correlation, berkowitz1968influence, annapureddy2016enhanced}. At room temperature, the mechanical properties such as hardness, yield, flow and tensile strengths decrease with the increase in grain size \cite{yuan2014dependence, ponpandian2002influence, matsumoto2002correlation}. The band gap of the nanoparticle samples decreases with the increase in grain size, which can be tuned by changing the annealing temperature \cite{hu2016stabilized}. Apart from these, the grain size allows us to modify the physical properties without altering the chemical composition of the sample \cite{reddy2012particle, kaushiga2022influence, kaarthik2023improvement}. There are numerous studies for quantifying the effect of grain size on physical properties of nanoparticles prepared for different applications \cite{jung201731, wang1995effect, graham2005transferable, reddy2012effect}.

The crystallite size can be determined using the X-ray diffraction analysis while the grain size can be obtained by analysing the microscopic images. It is usually determined using ImageJ software by choosing a limited number of samples. There are two ways to analyse the grain size: (a) draw the line of fixed length and calculate the number of particles along the line. After dividing the length of several particles, one can get the average grain size, (b) choose a certain area and find the number particles in the same which will provide the average grain area or size \cite{reddy2012effect, reddy2012particle, kaushiga2022influence, kaarthik2023improvement}. One has to repeat this by drawing multiple lines in different directions and choosing numerous areas in the given sample. All these measurements are as per the American society's standards for testing and materials (ASTM E112-13). The measure of grain size using the above technique is time consuming and requires a lot of manual effort. To avoid this, one has to automate the measurement and get the grain size directly from the instrument by post-processing the images taken with the microscope.

There have been a lot of attempts to automate the grain size analysis using computational methods \cite{banerjee2019automated, bankole2019grain, ghayour2016brief}. Many methods function on edge detection technique and counting the inhomogeneities over a specific area or length \cite{kumar2018particle, wu2016grain, akkoyun2022automated, peregrina2013automatic}. Artificial intelligence and convolution neural networks (CNN) have recently been utilized to find the grain size \cite{baggs2020automated, mishra2021estimation}. These methods have limitations and require high-performance computing facilities for obtaining automated grain size \cite{heilbronner2000automatic, meng2018automatic, canny1986computational, zhang1984fast}. Here, we propose a scheme to automate the grain size analysis that utilizes the auto-correlation function, which has been extensively utilized for many applications in optical domain such as extracting the mode information for the perturbed beams\cite{self1, self2, self3}. This function detects the edges at which the intensity gets diminished and assumes that the grains are independent of each other, i.e., the distribution of grains can be described using the Gaussian correlation function whose width is equal to the grain size.

For finding the grain size, we consider that the particles are randomly distributed which are similar to the inhomogeneities present on the rough surface. In this type of distribution, the particles are self correlated that can be described with Dirac-delta function. In reality, we can describe this using narrow function such as Gaussian distribution function with finite width. The width of this distribution function provides the average grain size \cite{self1}. The intensity auto-correlation function can be determined computationally for the experimentally obtained microscopic images using Matlab by using the following mathematical formula: 
\begin{equation}
    C = \frac{\ev{I_1(x_1,y_1)I_1(x_2,y_2)}}{\ev{I_1(x_1,y_1)}\ev{I_1(x_2,y_2)}}
\end{equation}
where $I_1(x_1,y_1)$ and $I_1(x_2,y_2)$ are intensities at two different spatial points of the same image. For random distribution, the intensity correlation strength is maximum if $(x_1,y_1)=(x_2,y_2)$, and it decreases with the increase in spatial separation. Here, we use the auto-correlation function for finding the grain size by utilizing the images obtained by Scanning Electron Microscope (SEM) and validate our results for recently prepared samples for utilizing to develop magnetic field sensors and energy harvesting devices. 

\begin{figure}[h]
   \includegraphics[width=8.75cm]{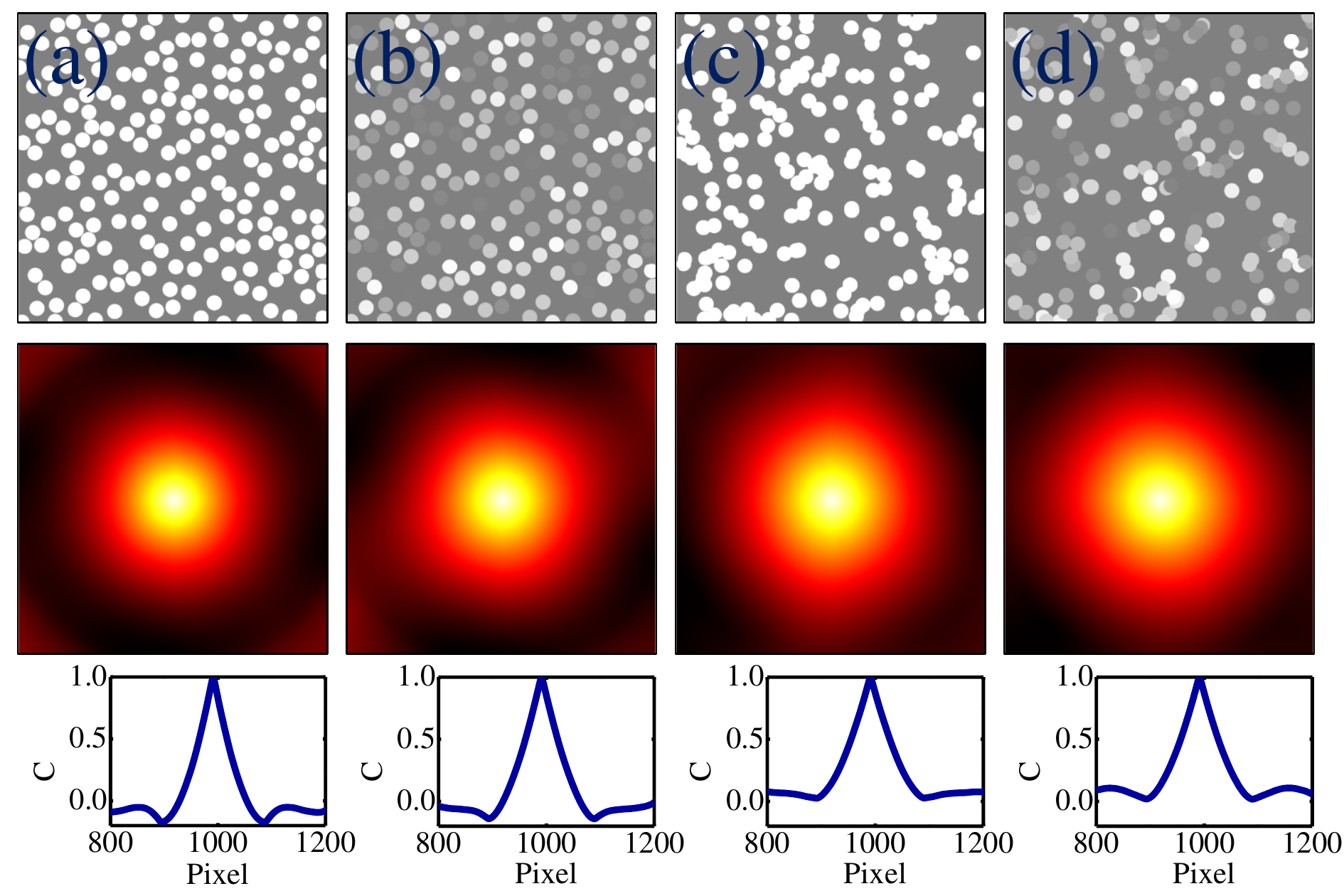}
    \caption{ (Colour Online) The different types of numerically simulated particle distributions (top), the corresponding intensity auto-correlation functions (middle) and their 1-D line profile plotted between correlation coefficient (C) vs pixel number (bottom).}
    \label{fig:expt}
\end{figure}

To verify our technique, first we proceed with the numerically simulated particles. For the same, we have considered the following four types of distributions for grains (a) non-overlapping particles with equal intensity, (b) non-overlapping particles with varying intensity, (c) overlapping particles with equal intensity, and (d) overlapping particles with varying intensity that covers all types of experimentally prepared samples. The distribution of simulated particles have shown in top of Fig. \ref{fig:expt}. We have simulated the circular particles of diameter 100 pixels and with total number of pixels as 2048$\times$2048. We have considered the total number of particles as 200. The position of the particles has been chosen randomly and added repeatedly based on the condition of overlapping/non-overlapping along with equal/unequal intensities. After obtaining the desired number of particles with proper distribution, we have used the auto-correlation function for finding the particle's size. The particle's size is determined by the width of the Gaussian function used for fitting the line-profile of the correlation function. The width is found for the Gaussian function at which its value is reduced by $1/e^2$ of its maximum value. 

Figure \ref{fig:expt} shows the simulated particles and their distribution (top), the corresponding auto-correlation function (middle) and its line profile (bottom). It is clear from the rows 2 and 3, that the obtained auto-correlation functions and their widths are same for all the four scenarios. It confirms that our technique works for all types of particle distributions as we have obtained the constant grain size of 100 pixels for both overlapping and non-overlapping particles as per the simulation. The auto-correlation is only dependent on the size of the particles. One can note that for all four cases, the grain is found as approximately 100 pixels, and it is the same for overlapping and non-overlapping particles. Although we have shown the results for circular particles, the proposed method can also work for elliptical particles for which we get varying widths along two mutually orthogonal directions and square particles.
\begin{figure}[h]
   \includegraphics[width=7cm]{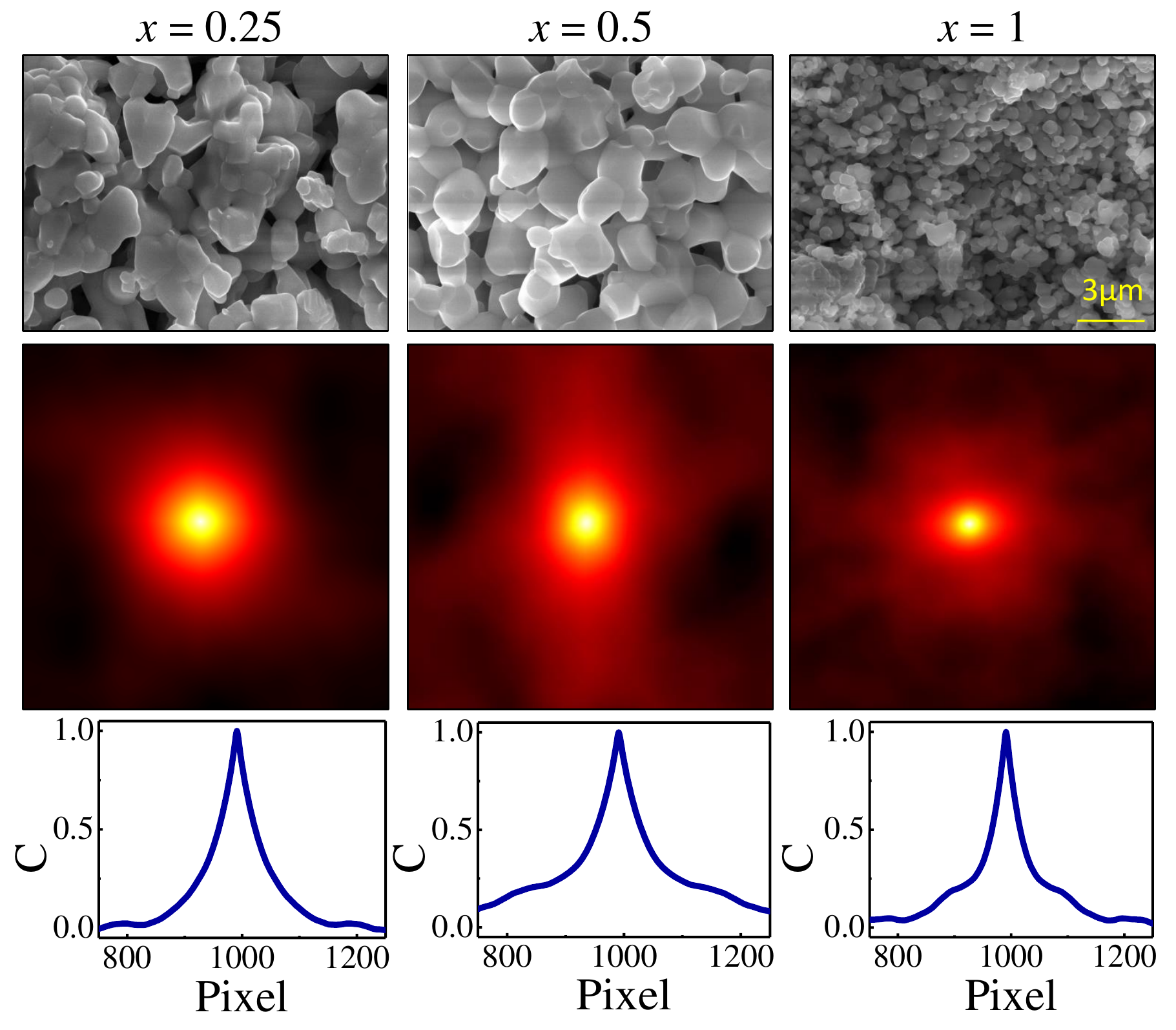}
    \caption{{The experimentally prepared particle distribution of Pr and Bi doped YIG (top) with the chemical formula of Pr$_x$BiY$_{2-x}$Fe$_5$O$_{12}$ where $x$ = 0.25 (left), 0.5 (middle) and 1 (right), the corresponding auto-correlation function (middle) and its line profile (bottom).}}
    \label{fig:expt1}
\end{figure}

Now, we proceed to validate our technique by verifying the grain sizes for experimentally prepared samples. We have prepared Pr and Bi-doped Yttrium Iron Garnet (YIG) nanoparticles with the composition of Pr$_x$BiY$_{2-x}$Fe$_5$O$_{12}$ by using sol-gel auto-combustion method for different values of $x$=0.25, 0.5 and 1. For the microscopic images, one has to find the pixel size by calibrating with respect to the provided scale at bottom of the image. By knowing the dimensions and total length of the sample, we obtain the pixel size just by dividing total length by total number of pixels. Further, we obtain the width of auto-correlation function in number of pixels that after multiplied with pixel size, gives the exact grain size.

Figure \ref{fig:expt1} shows the Pr and Bi-doped images for different concentrations of Pr $x$=0.25 (left), 0.5 (middle) and 1 (right) in top, the corresponding 2-D auto-correlation functions (middle row) and their line profiles (bottom). From the widths of the correlation functions, it is evident that the grain size decreases with the increase in concentration of Pr$_x$ which is also clear from the microscopic images. The ellipticity has been measured using the widths ($l_x, l_y$) of the correlation function corresponding to $x$, and $y$ directions with the help of the following relation,
\begin{equation}
    \epsilon = \frac{l_x-l_y}{l_x}.
\end{equation}

\begin{figure}[h]
   \includegraphics[width=7.5cm]{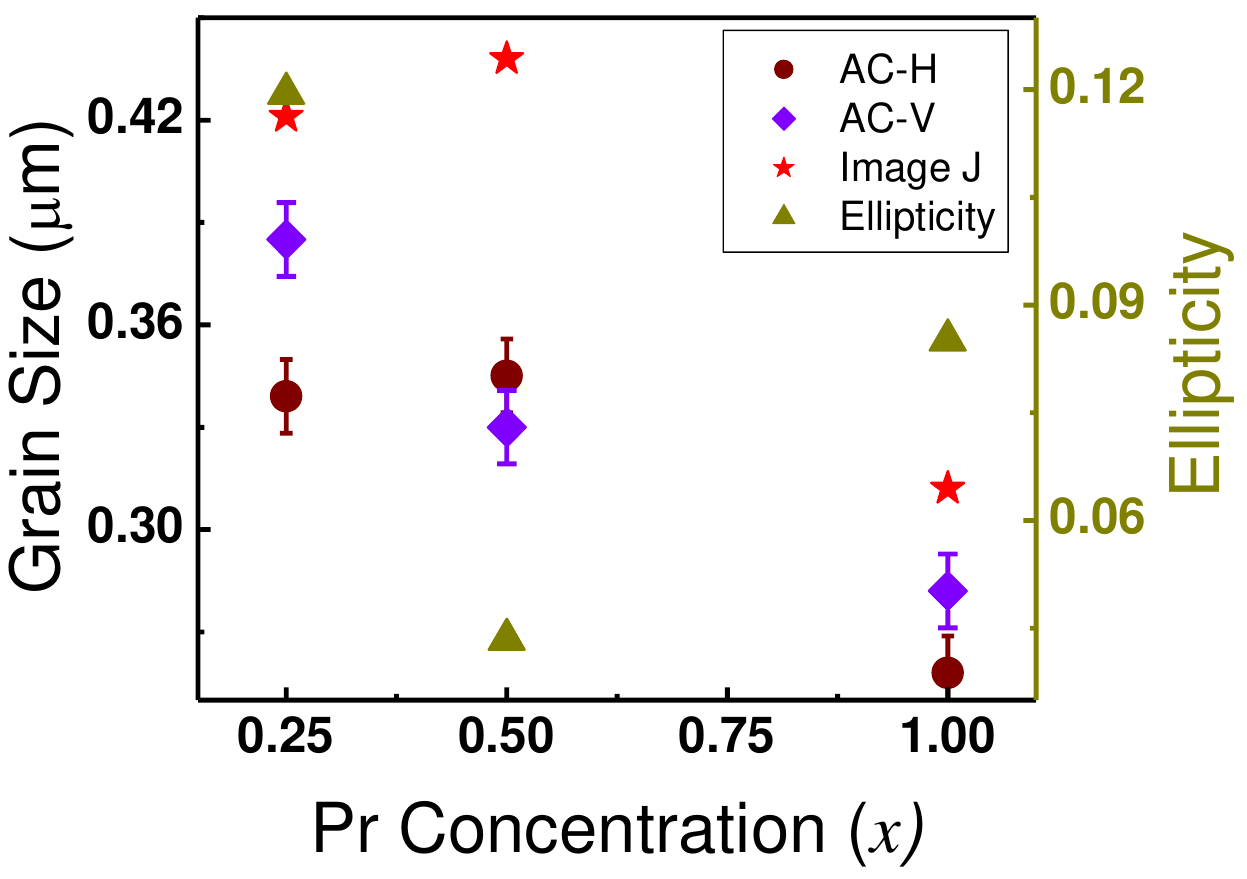}
    \caption{ (Colour Online) The variation of grain size measured along two orthogonal, horizontal (AC-H) vertical (AC-V), directions using autocorrelation function and manual technique using Image-J software along with the ellipticity of the particles as a function of Pr concentration $x$.}
    \label{fig:expt2}
\end{figure}

We further show the quantification of size and ellipticity of the grains in Fig. \ref{fig:expt2}. We have found the grain size along horizontal and vertical axes (AC-H, AC-V) by plotting the line profiles for 2-D autocorrelation function. For comparison, we also found the average grain size using the manual technique i.e. Image-J software. The grain size determined using autocorrelation function is approximately matching with manual measurements. The deviation may arise due to the fact that autocorrelation function takes the average over all the particles whereas manual techniques take the average over limited number of particles. The grain size is maximum for $x$=0.25 and minimum for $x$=1. The asymmetry is maximum for $x$=0.1, and the particles are more elliptical when compared to the other two samples. The error bars have been found after fitting the data with Gaussian profile. The results can be well matched by observing microscopic image with the naked eye.

\begin{figure}[h]
   \includegraphics[width=8.8cm]{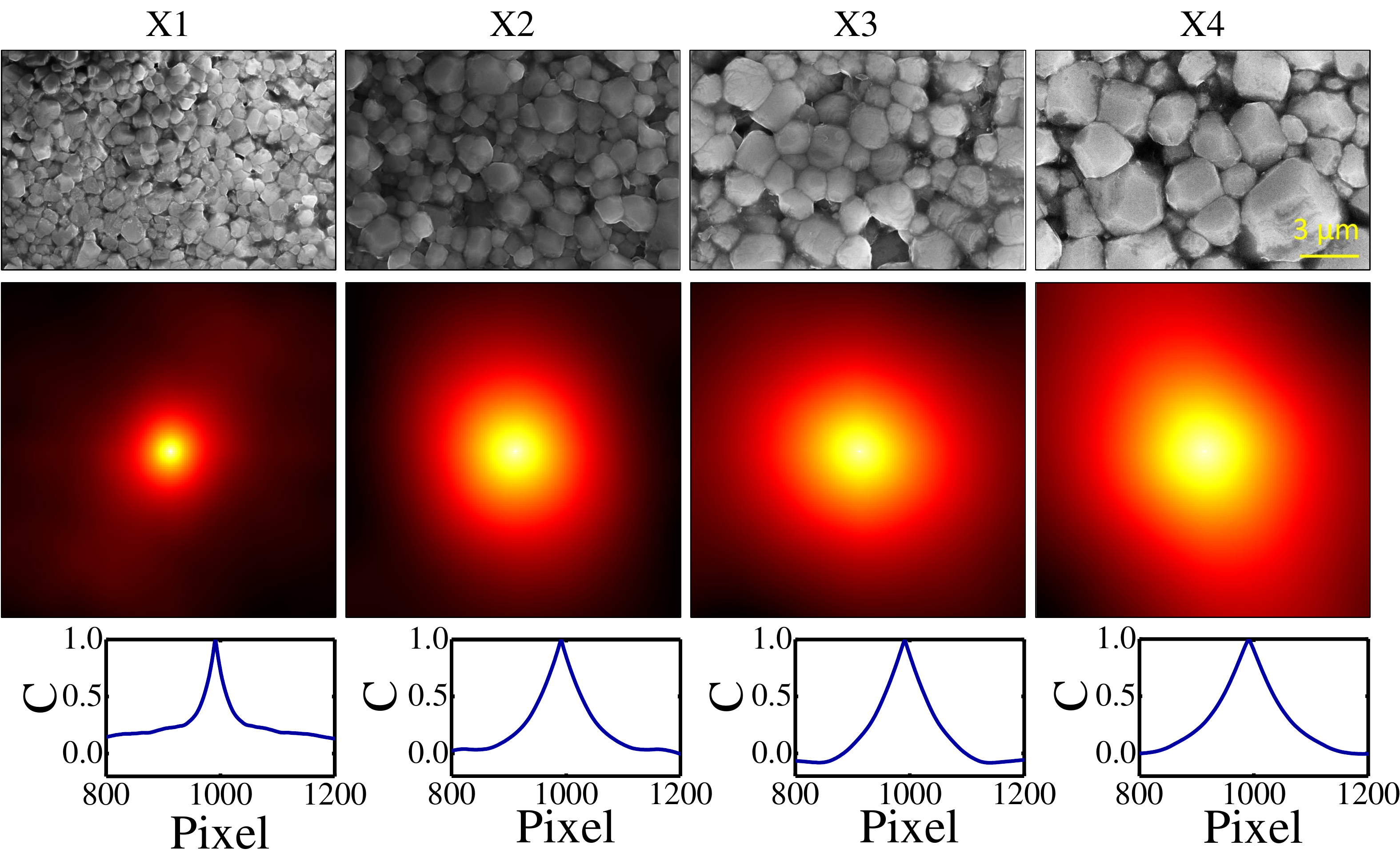}
    \caption{(Colour online) The microscopic images (top), of X1, X2, X3 and X4 as mentioned, the 2-D auto-correlation function (middle) and the corresponding line profiles (bottom).} 
    \label{fig:expt3}
\end{figure}

We also verify our technique for the samples prepared for energy storage devices \cite{kaushiga2022influence} using Ba, Sr, Bi and Ti oxides. We have prepared the four different samples, which are termed as X1, X2, X3 and X4 whose chemical compositions are as follows:
\begin{enumerate}
    \item X1 - Ba$_{0.6}$Sr$_{0.4}$TiO$_3$
    \item X2 - 99\% Ba$_{0.7}$Sr$_{0.3}$TiO$_3$ and 1\% BiMg$_{0.5}$Ti$_{0.5}$O$_3$
    \item X3 - 96\% Ba$_{0.7}$Sr$_{0.3}$TiO$_3$ and 4\% BiMg$_{0.5}$Ti$_{0.5}$O$_3$
    \item X4 - 95\% Ba$_{0.7}$Sr$_{0.3}$TiO$_3$ and 5\% BiMg$_{0.5}$Ti$_{0.5}$O$_3$.
\end{enumerate}

\begin{figure}[htb]
   \includegraphics[width=7.0cm]{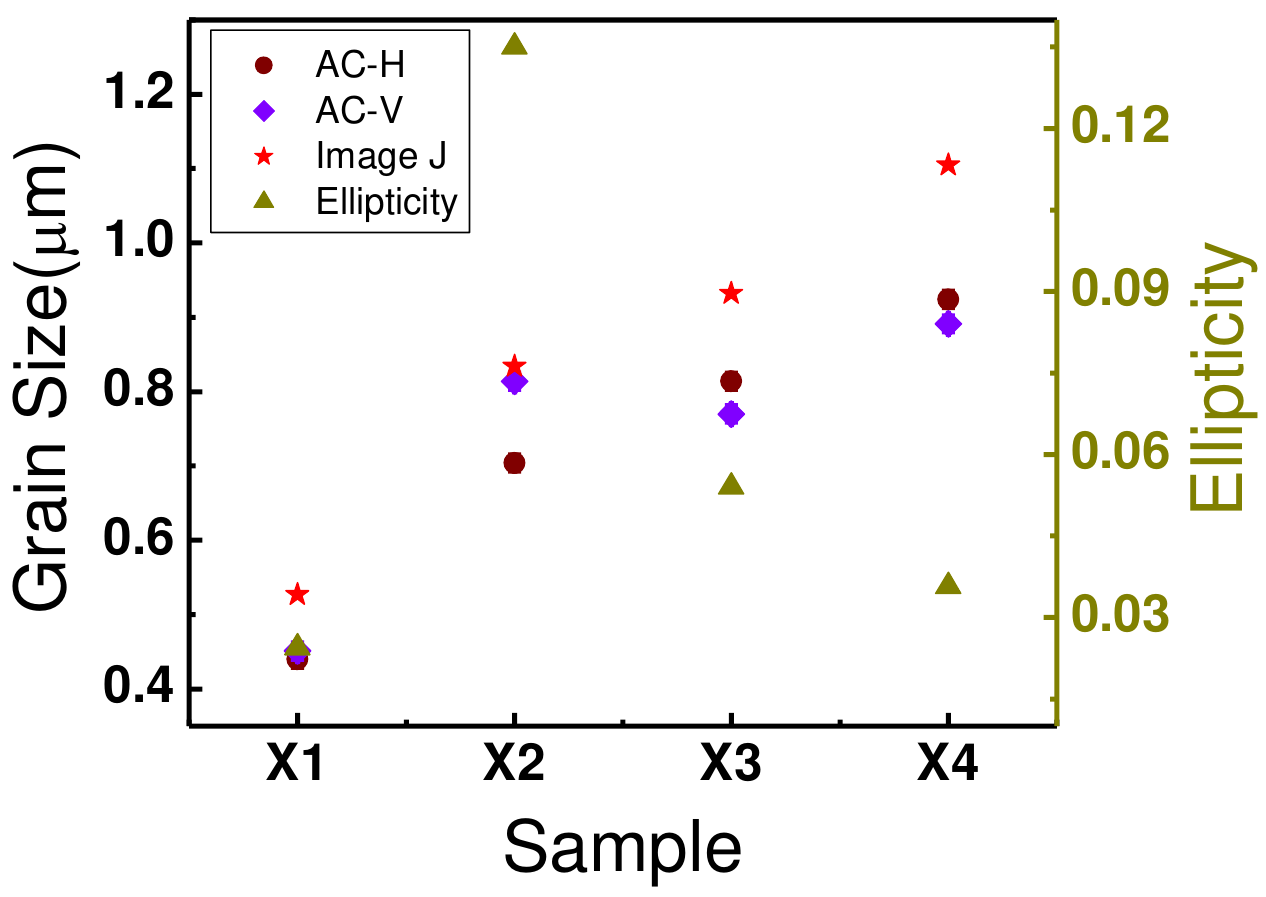}
    \caption{ (Colour Online) The grain size (top) and ellipticity of the nano-particles for the experimentally prepared samples X1, X2, X3, and X4.}
    \label{fig:expt4}
\end{figure}

The samples have been prepared using the chemical auto-combustion method. Figure \ref{fig:expt3} shows the microscopic images of four samples (top), the corresponding 2-D auto-correlation function (middle) and the line profile of the correlation function (bottom). We have arranged the samples as per the increasing grain size, and the particles are almost spherical in shape. The same has been confirmed using the ellipticity measurements.

Figure \ref{fig:expt4} shows the grain sizes measured using autocorrelation as well as manual techniques and the ellipticity for the four prepared samples. It can be observed that the grains are almost spherical, and ellipticities are obtained as less than 0.06. These results are in good agreement with the values calculated using manual methods. The errors are very small so that the error bars merge inside the point. With this, we conclude that the auto-correlation method works better for grain size analysis, as the change in intensity at the boundary is the key element for this method. 

In conclusion, we have proposed and verified experimentally a scheme based on the auto-correlation technique to measure the grain/particle size and its asymmetry. The auto-correlation function has been extensively used in the optical domain; however, here, we applied it for the material characterisation. We have analysed the grain size using this technique for both theoretically simulated and experimentally obtained particles. The results are in good agreement with the results obtained through manual techniques. This method can be utilized for all the samples for grain size analysis which can be implemented microscopy very easily using a simple Matlab or Python code.

SGR acknowledges the fund from SERB-DST through start-up research grant SRG/2019/000957. 

The authors declare no conflicts of interest related to this article. 

\bibliographystyle{tfnlm}
\bibliography{References}

\end{document}